\documentclass[10pt,journal,compsoc]{IEEEtran}

\ifCLASSOPTIONcompsoc
  \usepackage[nocompress]{cite}
\else
  \usepackage{cite}
\fi
\usepackage[hyphens]{url}
\usepackage{hyperref}
\usepackage[hyphenbreaks]{breakurl}

\usepackage{listings}

\usepackage{pgfplots}

\usepackage{tikz}
\usetikzlibrary{positioning,arrows,calc,shapes}
\usetikzlibrary{fit}
\usepgfplotslibrary{units}
\pgfplotsset{compat=newest}
\usetikzlibrary{decorations.pathmorphing}
\usetikzlibrary{decorations.markings}
\usetikzlibrary{shapes.geometric}
\tikzset{
    on grid,
    node distance=3cm,
    auto,
    myarrow/.style={
        draw=black,
        thick,
        ->,
        shorten <=3pt,
        shorten >=3pt,
    },
    mycircle/.style={
        draw=black,
        shape=circle,
        very thick,
        inner sep=3pt,
        inner ysep=5pt,
        text width=0.75cm,
        align=center,
        minimum size=0.75cm,
        rounded corners,
    },
    mytriangle/.style={
        draw=black,
        regular polygon,
        regular polygon sides=3,
        align=center,
        rounded corners,
        very thick,
        inner sep=3pt,
    },
    myrectangle/.style={
        draw=black,
        shape=rectangle,
        very thick,
        rounded corners,
        align=center,
        inner sep=7pt,
        inner ysep=7pt,
        text width=2.1cm,
        minimum size=0.5cm,
        minimum height=1.5cm,
        font=\footnotesize
    },
    mysquare/.style={
        draw=black,
        shape=rectangle,
        very thick,
        rounded corners,
        align=center,
        inner sep=7pt,
        inner ysep=7pt,
        font=\footnotesize
    },
    main node/.style={
        circle,
        align=center,
        draw,
        text width=.7cm,
        minimum size=.7cm,
        inner sep=7pt,
        font=\footnotesize
    },
    mythinsquare/.style={
        draw=black,
        shape=rectangle,
        rounded corners,
        align=center,
        inner sep=4pt,
        font=\footnotesize
    },
    process/.style = {
        draw,
        shape=rectangle,
        minimum height=3.5em,
        minimum width=3em,
        line width=1pt
    },
    unclocked/.style = {
        draw,
        shape=rectangle,
        minimum height=3.5em,
        minimum width=3em,
        line width=1pt,
        dashed
    },
    multiplexer/.style={
        draw,
        shape=trapezium,
        shape border uses incircle,
        shape border rotate=270,
        minimum size=20pt
    },
    ram/.style = {
        draw,
        shape=rectangle,
        minimum height=3em,
        minimum width=3em,
        line width=1pt
    }, toplevel/.style = {
        draw,
        shape=rectangle,
        minimum height=10em,
        minimum width=10em,
        line width=1pt,
        black!20,
        dashed
    },
    mux/.style = {
        draw,
        shape=rectangle,
        minimum height=1.5em,
        minimum width=1em,
        line width=1pt
    },
    empty/.style = {
        shape=rectangle,
        minimum height=3em,
        minimum width=3em
    },
    block/.style = {
        draw,
        shape=rectangle,
        minimum height=3em,
        minimum width=3em,
        line width=1pt
    },
    control/.style = {
        draw,
        shape=circle,
        minimum height=7em,
        minimum width=3em,
        line width=1pt
    },
    >=latex',
}

\usepackage{textcomp}

\begin{document}

\title{SME: A High Productivity FPGA Tool for Software Programmers}

\author{\IEEEauthorblockN{Carl-Johannes~Johnsen$^*$,
        Alberte~Thegler$^*$,
        Kenneth~Skovhede$^*$,
        and~Brian~Vinter$^\dagger$}\\[0.5em]
        \IEEEauthorblockA{\{$^*$Niels Bohr Institute, University of Copenhagen, $^\dagger$ Faculty of Technical Sciences, Aarhus University\}}}

\IEEEtitleabstractindextext{%
\begin{abstract}
For several decades, the CPU has been the standard model to use in the majority of computing. While the CPU does excel in some areas, heterogeneous computing, such as reconfigurable hardware, is showing increasing potential in areas like parallelization, performance, and power usage. This is especially prominent in problems favoring deep pipelining or tight latency requirements. However, due to the nature of these problems, they can be hard to program, at least for software developers.

Synchronous Message Exchange (SME) is a runtime environment that allows development, testing and verification of hardware designs for FPGA devices in C\#, with access to modern debugging and code features. The goal is to create a framework for software developers to easily implement systems for FPGA devices without having to obtain heavy hardware programming knowledge. This article presents a short introduction to the SME model as well as new updates to SME. Lastly, a selection of student projects and examples will be presented in order to show how it is possible to create quite complex structures in SME, even by students with no hardware experience.
\end{abstract}

}

\maketitle

\IEEEraisesectionheading{\section{Introduction}\label{sec:introduction}}
The use of CPUs for programming is an incredibly flexible approach that allows the same hardware to be deployed in many different scenarios, and also enables seamless updates to functionality. The basis of a CPU is the sequential programming model where one instruction is executed at a time, resulting in a versatile state machine. With evolving programming languages, it is possible to produce increasingly complex solutions with very little understanding of the underlying mechanics.

For this reason it seems attractive to attempt to apply this model when developing reconfigurable hardware. The prospect of being able to tap into the vast pool of programmers and existing applications would clearly be an desirable goal.

However, directly applying the sequential programming model to hardware designs would require a long continuous circuit signal, or a complex state machine mimicking sequential behaviour. This would significantly reduce the achieved speed and area utilization of an FPGA.

To better utilize the ability to perform operations in parallel, it is necessary to either infer parallelism from sequential code or design the program with a high degree of parallelism. For the former approach, work on extracting parallelism has been an ongoing research field for decades with few promising results~\cite{levine1991comparative}. For the latter approach, research shows that parallel systems are harder for programmers to work with~\cite{szafron1996experiment}.

To balance these two extremes, we created SME which is a hardware design model that builds on the formal algebra of CSP~\cite{hoare1978communicating}. What CSP does well, is expressing concurrent programs without introducing the hazards normally found in a parallel programming model. Teaching CSP as a programming model to students shows the model itself is simpler than other methods~\cite{vinter2019teaching} and initial results also show that the SME model allows students familiar with sequential programming to pick up the parallelism~\cite{marchant2019teaching}.

\subsection{Contribution}
In this paper, we will give an overview of our programming model, Synchronous Message Exchange (SME), which we believe is a useful tool for FPGA development. SME is a CSP-derived programming model, that is used to describe hardware and can be translated into VHDL. Because SME is built on CSP, we can avoid deadlocks and race conditions, which are normally associated with concurrent programming.

We will start by introducing the concepts of SME, in Section~\ref{sec:sme}, followed by the updates SME has received to the back-end in the past few years, in Section~\ref{sec:back}. Finally, we'll present some examples of the capabilities of SME, in Section~\ref{sec:ex}.

\subsection{Related work}
In this section, we will give a short introduction to a number of existing solutions seeking to simplify the programming of FPGAs.

\subsubsection{Hardware Description Language}
FPGAs are usually programmed using a Hardware Description Language (HDL). One such language is Very High Speed Integrated Circuit (VHSIC) HDL, commonly referred to as VHDL. While VHDL can be programmed sequentially, it is intended primarily to be a parallel programming language. As such, in order to gain increased performance, the implementation must be restructured into a concurrent/parallel structure. This however, can be a very tedious task.

Despite the continued evolution of the language, vendors are not implementing recent updates to VHDL and so it is becoming stagnant, from a users perspective. Xilinx have adopted VHDL-2008 in 2019~\cite{xilinx-vhdl2008} and Intel in 2014~\cite{intel-vhdl2008}, with both only having limited support for the standard to this day. This means that even though there have been updates to the language, there is no direct way of using these updates without vendor implementation.

\subsubsection{High-level Synthesis (HLS)}\label{sec:rw-hls}
HLS~\cite{xilinx-hls, intel-hls} is one of the most popular approaches to describe hardware using a higher level language. It is implemented using either C or C++. This gives a familiar approach for developers, as many have used a C based syntax before. It also allows for rapid development, as the tools handle a lot of the techniques, such as control flow and pipelining, required for implementing functionally correct hardware.

However, C is a sequential programming model designed for CPUs, rather than FPGAs. As such, it is not trivial to gain performance, as it relies on automatic derivation of parallelism. Furthermore, it introduces significant overhead in exchange for it being safe and correct. An example of why automation introduces overhead can be seen in the general way it handles data hazards: It locks everything. Locking ensures that the hardware is both safe and correct, but at the cost of a performance penalty.

HLS does provide a solution to these problems in the form of the  \texttt{pragma} decorator, similar to using \texttt{OpenMP}~\cite{openmp} for parallel programming. These \texttt{pragma} guide the compiler, giving it additional information to work with, such as pointing out fake data hazards. While this allows for better results, it removes some of the original intent of HLS, as a user now has to alter their program and write it in a specific manner that differs from the original sequential model.

\subsubsection{OpenCL}\label{rw-opencl}
Another popular approach is using OpenCL~\cite{xilinx-opencl, intel-opencl}. As with HLS, the intent is to attract GPU programmers. This seems like a better idea than HLS, given that OpenCL is a parallel programming model. It is also a great fit for writing accelerator solutions, given the host device programming model of OpenCL. This host device model ensures that the programmer does not have to handle all of the fine-grained logistics of the implementation, such as moving data to and from external memory.

Intel is moving to the newer programming model SyCL~\cite{intel-sycl}, introduced by the Khronos group. This has the same theoretical underpinning as OpenCL, but is written in C++ and is single source, as in both host and device code is described in a single source file. This allows for device agnostic host development.

A problem with the OpenCL programming model, is that the vendor manuals states that it should be written using only one thread, which contradicts the proposal of using parallel programming. If multiple threads are used, and the scaled design does not fit onto the FPGA, the OpenCL compiler would have to reuse the hardware components in order to complete the computations. This would introduce additional overhead and therefore it is not always optimal to use multiple threads in OpenCL.

\subsubsection{Chisel}\label{sec:rw-chisel}
Chisel~\cite{chisel} is a programming model for constructing hardware, which is written in Scala. It is made by the same team that made RISC-V~\cite{riscv} at Berkeley. The team behind Chisel provides regular updates and report that it has good use in education.

The Chisel approach is similar to the the SME approach in that items are created as isolated units with explicit communication. While Chisel is described as an embedded language it is actually a library in Scala. This difference has a minor impact on the syntax where the programmer has to use \texttt{when(condition)} but can also use \texttt{if(condition)}. Apart from the syntactic difference, this has the implication that the programmer needs to be able to keep two states of the program in mind, one that is running in Scala, evaluating the \texttt{if} statements, and another running the generated code, not being able to see the \texttt{if} statements. It also means that the code is generated and cannot be debugged as the hardware code is generated by observing traces of the library execution.

\subsubsection{C$\lambda$ash}\label{sec:rw-clash}
C$\lambda$ash~\cite{clash} is a functional programming model, written in Haskell. It leverages the parallelism, which can be expressed in functional languages. As functional programs are naturally free from side effects, it is possible to extract a higher degree of parallelism than what is possible with imperative sequential programs.

As Haskell is a functional programming language it is common to solve problems with recursive function definitions. This common programming construct has limited support in C$\lambda$ash as it needs to unroll the recursion to fit statically allocated hardware. This is equivalent to statically inferring the number of repetitions in a loop, which degrades to the halting problem.

\section{Synchronous Message Exchange}\label{sec:sme}
Synchronous Message Exchange (SME) is a programming model, targeting FPGA development using a high-level language. The idea for SME was developed after a student project focused on CSP as a programming model for hardware design~\cite{pycsp-processor}. This project showed that due to the requirement of global synchronization, broadcasting, and latches, the amount of processes and channels would quickly explode when using CSP. Therefore, it was not going to be feasible as a standalone programming model. During the project, the isolation of having share-nothing processes proved very useful when programming hardware models. As a result, SME was created by implementing the usable elements of CSP and leaving out the elements that did not fit into a programming model for hardware. Elements like external choice was omitted but the share-nothing property was introduced into SME.

An SME program is structured much like a CSP program. It has sequential processes, which share nothing, except for their means of communication. In CSP, these means are rendezvous channels, where the communication takes place once both ends of the channel are "ready". This is different in SME, which uses broadcast buses. Here, values are broadcast from a writing process to one or more reader processes. This broadcasting mechanism is driven by a hidden global clock, rather than the rendezvous strategy in CSP.

Given that SME is targeting hardware development, an SME simulation revolves around the hidden clock. During each clock cycle, every process is triggered exactly once. The order in which the processes are triggered, depends on their triggering strategy and their connection in the network. From this trigger mechanism, we build a dependency graph, from which we create an Abstract Syntax Tree (AST) based on the runtime instances of processes and buses. This AST is used for translating the model into VHDL, through a series of VHDL templates.

Given that SME is programmed using a high-level language, some low-level constructs, patterns or data manipulations can be hard to express. We try to counter this by aiming for the code to be as close to the input source code as possible. The idea is that low-level, platform specific optimizations can be made in the low-level language, even by a platform expert, since the generated code is human-readable.

We will now explain the base concepts of the SME model. It will explain these concepts from a software developers point of view, through similarity with CSP, and how they relate to hardware design concepts, through similarity with VHDL.

\subsection{Processes}\label{sec:sme-processes}
A process in SME is a sequential block of code, with each process being isolated from the other processes. Like in CSP, processes in SME share nothing except for buses and constants.
There are two types of processes that exists in SME: simple processes and simulation processes.

\subsubsection{Simple Processes}\label{sec:sme-simple-procs}
This type of process is, as the name states, the simplest form of SME processes.
The defining characteristics of a simple process are:
\begin{itemize}
    \item \textbf{An \texttt{OnTrigger} function} -
        which contains a set of statements dictating what the process will do every clock cycle.
        These statements must be static, or at least statically bounded, such as the case with loops.

        In every clock cycle, all statements within the \texttt{OnTrigger} function will be completed. This also means that the developer needs to consider how many operations an \texttt{OnTrigger} should handle. More operations will have a negative impact on the length of a circuit path, increasing the required time span of the clock cycle.
    \item \textbf{An optional set of input buses} -
        from which the process can read, but cannot write to.
        Values read from a bus will not change during a clock cycle.
    \item \textbf{An optional set of output buses} -
        to which the process can write, but cannot read from.
        Only the last values written within a clock cycle will be propagated on the bus.
    \item \textbf{An optional set of variables} -
        which, like the statements in the \texttt{OnTrigger} function, are limited to static data structures.
        For example, a process cannot have lists, as they are dynamically allocated, but it can have arrays, as these can be statically allocated.
        Unlike the buses, the variables can be read and written multiple times within the same clock cycle. Each variable is isolated within the process and, as such, is inaccessible outside the process.
        If a variable does not change over the entire simulation, it will be translated into a constant, which will be much more efficient. Simulations are described further in Section~\ref{sec:sme-simulation}.
    \item \textbf{An optional set of functions} -
        which are either pure and without side-effects, or are limited to only have an effect on either the processes buses or internal variables.
        The functions are, like the variables, isolated within the process and cannot be called outside of the process.
    \item \textbf{An optional \texttt{clocked} attribute} -
        which indicate the triggering strategy of the process, which is described more in depth in Section~\ref{sec:sme-simulation}.
        If the process has the \texttt{clocked} attribute, it is called a clocked process and will be triggered by the hidden clock. If it does not have the clocked attribute it is an unclocked process and is only triggered once all the processes on the other end of its input buses have executed.
        Regardless of the strategy, a process will be triggered exactly once per clock cycle.
    \item \textbf{An optional \texttt{ignore} attribute} -
        which indicate whether or not the process will be translated into the hardware model.
        If the process has the ignore attribute, it wont be translated and will only exist during the SME simulation.
        Because of this, the static limitation on data structures and statements are lifted.
\end{itemize}

Processes in CSP and simple SME processes are much alike.
Both are sequential and share nothing except for their means of communication. The key difference, is how they are triggered. A CSP process runs immediately, and only once, where a simple process in SME only runs when triggered, which happens exactly once per clock cycle. However, since SME can be implemented in CSP~\cite{thegler2018cspm}, the same behaviour can be achieved in CSP. While CSP processes can perform context switches based on both timing and communication, SME simple processes does not. A simple process also does not, at least not directly, feature the CSP \texttt{ALT}, as this would require waiting for a branch.

Since simple processes can be translated into VHDL processes, they are also very alike. A simple process can be compared with a VHDL process, where the buses are ports, with a reset signal and either a ready signal or the clock signal as sensitivity signals. The generated VHDL processes will feature reset logic, which resets the variables and the output buses, followed by the statements in the \texttt{OnTrigger} function.

\subsubsection{Simulation Processes}\label{sec:sme-sim-procs}
Simulation processes share the same defining characteristics as a simple process. Simulation processes has the \texttt{clocked} and \texttt{ignore} attributes set by default. As such, they are not constrained to static constructs, like the simple processes, and can utilize every part of its implementation language. They are usually used to drive the input and verify the output of the SME simulation. A simulation process will trigger only once during a simulation, so the trigger function is spread over multiple clock cycles. This is leveraged by utilizing an asynchronous programming model, which allows waiting for the clock to tick.

\subsection{Buses}\label{sec:sme-buses}
Buses are the means of communication between processes. They contain a collection of fields, which are bundled together by a user, based on context.
For example, it makes sense to bundle all of the signals that go into a port of a RAM, as they relate to each other.

Buses are directly comparable to wires, registers, and latches in hardware, which are limited to static constructs, and therefore the allowed data type of the fields in buses are also limited to static constructs.

Buses are broadcast between the connected processes.
There can be multiple reading and writing processes to a bus, but only one writing process for each field in a bus. This is as there cannot be double drivers to wires in hardware, at least not without some form of conflict resolution.
If no process writes to a field in a bus during a clock cycle, the last written value will be latched for the following clock cycle.
Buses can, like processes, be clocked or unclocked.
Clocked buses won't propagate their value until the following clock cycle.
Unclocked buses will propagate their value to the subsequent processes within a clock cycle.
All buses, that a clocked process reads from, must have an initial value or be guarded by some field which have an initial value.
If a clocked process attempts to read a field that does not have a value, it will receive an indeterminate value. SME handles this by throwing an exception, indicating that the developer has to explicitly handle this.
This is controlled with the \texttt{InitialValue} attribute on fields, or the \texttt{InitialisedBus} attribute on buses, which set all of the fields to have the default datatype value as initial value.

\subsection{Network}\label{sec:sme-network}
As mentioned before, an SME simulation consists of processes and buses, which is collectively known as a network. A network describes how the processes and buses are instantiated and connected. As the simulation has not been started when a network is being created, a user can use both static and dynamic constructs. As such, dynamic collections and strategies can be used for instantiation. For example, to define a pipeline of a single type of process, one can utilize lists and loops. It is during the instantiation that every process, bus and how they are connected, is registered by SME.

There are multiple strategies for instantiating a network. We propose that every process instantiates its own output buses, given that there can only be one writing process per field in a bus and this way, all buses needed in the network will be instantiated correctly. This might not always work, such as the case where multiple processes write to individual fields of a bus, in which there should not be multiple instances of that bus. Therefore, this strategy should not be considered a catch-all solution for all problems, but is generally sufficient. On a related note, zombie instances of buses are allowed, as they won't be translated into hardware, but these are not recommended.

\subsection{Simulation}\label{sec:sme-simulation}
Once the network has been set up, SME creates a dependency graph, where processes are nodes and buses are edges.
This dependency graph indicates how the processes are connected and the order that they should be triggered in.
In this dependency graph, there cannot be cycles of unclocked processes and buses, as it would result in a short circuit on the board, when the network is translated into hardware.

Processes are triggered based on two strategies: through the hidden clock, which controls the clocked processes, or through whether or not all of the input buses have been written to, which controls the unclocked processes.
Processes are independent either if they are clocked processes or if they do not share opposite ends of a bus instance. Note: chains of processes connected by bus instances are not considered independent.

Since each process is sequential, the parallelism in SME is explicit. In order for something to run in parallel, a user needs to solve the problem by using multiple clocked processes.
Every independent process is triggered concurrently.
If two unclocked processes depend on each other through a bus instance, the process with the bus instance as input wont be triggered until the other process has been triggered and completed executing.

After the dependency graph has been created and verified, the simulation can begin.
A simulation is a run of multiple clock cycles.
The algorithm for driving the simulation is as follows:
\begin{enumerate}
    \item Propagate all of the \texttt{clocked} buses.
    \item Collect all of the \texttt{clocked} processes into an \textit{active} set.
    \item Trigger all of the processes in the \textit{active} set.
    \item Propagate all of the \texttt{unclocked} buses, which are edges out of the nodes corresponding to the processes in the \textit{active} set.
    \item All of the processes, whose input edges in the dependency graph have been propagated, become the new \textit{active} set.
    \item Repeat steps 3 - 5 until there are no more processes, which have not been triggered.
    \item Check if the simulation stop condition has been met
        \begin{itemize}
            \item If the stopping condition has been met, the simulation terminates.
            \item If the stopping condition has not been met, the simulation repeats from step 1.
        \end{itemize}
\end{enumerate}

\subsection{Testing}\label{sec:sme-testing}
To verify whether a network is correct, the programmer has to write tests. This is done through simulation processes, which will either drive the network
with input, verify the output or both. As mentioned in Section~\ref{sec:sme-sim-procs}, a simulation process is not
limited to the synthesizable part of SME, but the developer can use the entirety of the front-end language, which is one of the great strengths
of SME.

For example, to produce an implementation that computes an image, it is not necessary to write a function to read the image. The image can simply be loaded through an external library and during each clock cycle, the pixel values of the image can be fed into the network on the buses.

During simulation, every value on every bus is recorded for every clock cycle. These values are then saved in a trace file, which is used to test the
resulting VHDL.

SME creates a VHDL test bench, loads this trace file and drives all of the input buses with values from the trace file. It then verifies that all of the values on all of the output buses matches the values from the
trace file. This helps in verifying that the resulting VHDL is clock cycle accurate, compared to the SME simulation.

This means that it is not necessary for the developer to create a test bench manually in VHDL, which can be very tedious. This ensures that testing the network can be done faster, compared to what would otherwise be possible, and it also makes the development simpler when using SME.

\subsubsection{GHDL}\label{sec:sme-ghdl}
The vendor tools for FPGAs are computationally heavy and take up a large amount of disk space of at least 30 GB. As such, it is not feasible to utilize these in continuous integration systems, which usually run in containers connected to lightweight resources.

We have used GHDL~\cite{ghdl}, which is a VHDL simulator using a code generation back-end, such as GCC. This is resulting in a fast simulation that runs on CPUs. By using GHDL, users of SME can get a quick verification on whether the VHDL model is correct, without using the heavy vendor tools

\subsection{Components}\label{sec:sme-components}
It is often useful to have optimized implementations of common constructs. In the software world this is handled by libraries, and in the hardware world it is handled by IP Cores. SME uses a mix of these two approaches called components. Components provide a high-level C\# library during simulation, which are then translated into IP Cores in VHDL. By using the components, the user does not have to write to the exact specification a vendor requires.

\section{SME updates}\label{sec:back}
SME has been introduced before~\cite{vinter2014synchronous,vinter2015bus,skovhede2016building}, but since last publication, a lot of changes have been made to the back-end of SME. While these do not have a direct impact on the usage of SME, they should help to make SME more future-proof and increase its performance.

\subsection{C\#}\label{sec:back-cs}
\subsubsection{.NET core}\label{sec:back-cs-dotnet}
Previous versions of SME was compiled using different compilers depending on the platform. On Microsoft Windows, the .NET Framework compiler was used, on Linux and OSX, the Mono project was used. Since then, Microsoft have purchased the Mono project and released a cross platform compiler, called .NET Core. Microsoft intends to have a unified .NET ecosystem, having the same compiler for every platform, eliminating the big performance gaps between the different compilers.

Another benefit of using the most popular compiler, is that new language features are no longer delayed on the non-Microsoft platforms. These are not guaranteed to be translatable into hardware, as they might introduce unknown patterns in the syntax trees. They can be freely used for the simulation processes and for setting up the network, as there are no restrictions here.

With the introduction of .NET Core 3.0, async Tasks now run in parallel. Since SME uses Tasks for running the dependency graph, they now also run in parallel during simulation. All hazards, that are introduced by the parallel execution, are handled by .NET core, however this is done so in an aggressive manner. It finds false dependencies on the buses, which results in some of the processes running in sequential order. While this does not break simulation, as everything is still handled concurrently, it decreases the runtime performance of the simulation.

\subsubsection{Roslyn}\label{sec:back-cs-codeanalysis}
In the first iterations of SME, the model was derived from the compiled binaries produced by a C\# compiler, such as Mono~\cite{monoproject} or Microsoft .NET~\cite{microsoft-net}. This was done by decompiling the intermediate language in the binaries into an Abstract Syntax Tree (AST). While this allowed SME to utilize optimizations made by the compiler, it also introduced unwanted constructs through these optimizations, such as variable names, since these optimizations are targeting CPUs, rather than FPGAs.

At some point, new instructions were introduced to the Common Intermediate Language (CIL)~\cite{cil} by the .NET compiler. These instructions were not recognized by the decompiler, resulting in SME not working.
While this could be circumvented by keeping all of the decompilation libraries up to date, this wouldn't be the best solution, as these libraries also introduced breaking changes to their API. This would then require substantial work on the back-end of SME.

Recently, Microsoft have begun releasing its compiler as an open source project called Roslyn~\cite{microsoft-opensource, roslyn}. Though not all parts are released yet, it allows developers to access the internal syntax trees and semantic models of a compiled .NET program.

By exchanging the decompiler with a compiler,
every node of the AST can now be linked to actual nodes in the compiled syntax tree. This both improves naming, as the original names can be obtained, and error messages, as these can now be linked directly to a specific location in the source code.

Currently, execution is slower, compared to the decompilation, due to the way the syntax trees and semantic models are build using Roslyn. Every time a node in the tree is exchanged, the entire compilation has to be recompiled.

One problem with using the compiler is the comparison of runtime types and compilation types. They might resemble the same type, but because they are in different compilation contexts, they are not directly comparable. The current solution is a naive string comparison of the fully qualified type name. This might be sufficient, but it is inherently a bad idea to compare objects, using only their string representation.

\subsection{Dynamic instantiation}\label{sec:back-dyn}
As described in Section~\ref{sec:sme-network}, an SME network is described by instantiating processes and buses. As this is done before running the simulation, there are no constraints on the constructs used for this, giving the option to dynamically build the network.

This wasn't always the case. Previous versions of SME relied on the type definitions, rather than the instances. For building the network, SME would gather all definitions of processes and buses and instantiate exactly one of each. Connections were generated based on the types, as there could only exist one instance in a network. The idea was that connections would then be created automatically instead of explicitly.

This became a problem for bigger designs featuring either multiple instances of the same process, such as registers in a pipeline, or design utilizing compositionality, where processes and buses are grouped together. In these cases, the user had to define the same process multiple times. This was partly solved by using templates for generating code, but it proved hard to debug from the template level.

Our new approach is to have the user construct the network explicitly, by manually instantiating and connecting everything. It might seem tedious, but it opens up the possibility of using dynamic structures and constructs, such as lists or loops, for describing the network. This is possible as it is not crucial how the entities are instantiated, as long as it is done before the simulation is started, at which point all structures has to be static. This new way of instantiation translates well to VHDL.

\subsection{SME Intermediate Language}\label{sec:back-smeil}
Different programming languages have different strengths which could prove useful in different situations. To better support multiple implementation languages, an intermediate representation for SME networks was developed~\cite{asheim2019smeil}. With an IL representation it would be possible to develop libraries in one language and invoke them from another. It also simplifies the generation of the resulting HDL as only a well defined set of constructs need to be supported. Due to resource constraints, the SMEIL approach is not fully developed.

\subsection{Formal verification}\label{sec:back-verification}
To verify that the created hardware model behaves as expected, test benches are usually the main method of testing. These test benches are often sufficient, but in some cases it is not enough to find critical errors in the hardware design. With formal verification it might be possible to locate some of these errors that have not been detected with the standard test bench.  While formal verification is the desired approach, it is also difficult and is not possible in all formats and languages. Therefore, it is not typically a standard tool for testing.

CSP~\cite{hoare1978communicating} is a process algebra that provides a method to express interaction between processes of concurrent systems. Some properties can be formally verified in systems described in CSP. This means that if hardware can be described in CSP, it will be possible to formally verify certain properties within the hardware that cannot, or is difficult to, be tested through the test bench. As previously mentioned in Section \ref{sec:introduction}, SME is derived from CSP and therefore for each SME model there will be an equivalent CSP model. This means that the SME model can be formally verified using the equivalent CSP model as it is directly translatable.\\

The system TAPS (Towards Automatic Program Specification)~\cite{thegler2018cspm, albertesspeciale} provides translation from SMEIL, see Section ~\ref{sec:back-smeil},  to the machine-readable version of the CSP process algebra, $CSP_M$~\cite{Scattergood1998}. The generated $CSP_M$ code will not only be equivalent to the SMEIL code from which it originates, but it will also include assertion statements to formally verify properties within the defined hardware model. These properties can be verified with the $CSP_M$ refinement checker tool, FDR4 (Failures-Divergences Refinement)~\cite{gibson2014fdr3}. Instead of having to create advanced test benches, TAPS provides a simple way to verify the hardware model via FDR4s assertion functionalities. FDR4 can, using TAPS, assert that the observed values of a channel, in a simulated SMEIL program, are in fact the only possible values communicated on that specific channel within a specified amount of clock cycles. The amount of clock cycles are defined, by the user, when simulating the system.

For example, a hardware design can have a 4 bit connection from one process to another, which can represent the number 0 through 15.
But the system might not have been designed for numbers above 10 to be communicated between the processes. It would, of course, be possible to test the network for numbers above 10, but for more elaborate designs it might be too complex for the developer to grasp. A formal verification with TAPS would be able to verify if it is possible for a number above 10 to be communicated between these processes with the current design.

The SME simulation observes the values throughout the simulation, then adds the values to the code and outputs it. It also defines the size of the range of the observed values, for example 4 bits for the range 0 - 15. These ranges are used by TAPS to create assertion statements for the verification.

Utilising the formal verification properties of the FDR4 refinement tool, the assertion statements of the translated $CSP_M$ code can be verified.

The SME model is simulated for a specified amount of clock cycles. It is necessary to simulate the system in order to get data to verify, but this also means that if the simulation has not been running long enough, the verification might not be adequate. This challenge has been discussed in~\cite{albertesspeciale}.

In order to be able to simulate the internal state between clock cycles, it is necessary for TAPS to simulate recursive processes and a possibility to retain the state between the cycles. When creating recursive processes in $CSP_M$ for formal verification, it is necessary to have the recursion stop eventually, since it is not possible to formally verify endless loops. Because CSP is not inherently synchronous, it is necessary to have the functionality to translate the clock that drives the SME circuit to the $CSP_M$ translation. Therefore TAPS provides the possibility of modeling a globally synchronous clock in $CSP_M$ that drives the network and terminates after a specific number of clock cycles.

\subsection{Student projects}\label{sec:stud}
Throughout the last couple of years, several different projects have been created with SME as a main structure. Each of the mentioned projects, in this section, have been performed by masters students at the University of Copenhagen with no prior knowledge of SME or FPGAs. These projects showcase the capabilities of SME, in that SME can be used for implementing hardware without having any prior knowledge or understanding of hardware programming. Some of them also show that the granularity of SME proves a challenge.

\subsubsection{Median filter}\label{sec:stud-sorting}
This project~\cite{troels-ynddal} was part of a pipeline, processing a stream of data coming from an X-ray line scanner for food inspection. The part in focus was a median filter.

It was implemented using a series of Block RAM (BRAM), which read out a window of pixels from the image. These pixel values were then sorted by a sorting network to find the median. Finally, the median was written to another BRAM holding the results. All of this was implemented by using SME.

The hardest part proved to be handling the transactions to and from the BRAM, as there is a latency when using BRAM. Each request to/from a BRAM requires at least 1 clock cycle, where the request is considered to be in-flight, before the request is handled. Upon reading, another clock cycle is introduced, before the response is ready.

The resulting implementation kept up with the performance requirements of the rest of the pipeline, such as the data rate from the line scanner.

\subsubsection{TCP controller}\label{sec:stud-tcp}
This project~\cite{janmark} worked on a hardware implementation of a TCP controller. The idea was to provide an open source interface for other hardware solutions, wishing to communicate with external entities.

Due to the steps of processing a TCP packet, where each layer of the TCP stack had to be processed in order of arrival, the solution had to hold a state. This state became more and more complex, once the controller had to handle other features of the protocol, such as partial packets and out of order packets.

While the project did end up having proposed an architecture for handling basic TCP communication, the dynamic nature of network communication was shown to be more fitting for a higher level abstraction, than the one provided by SME.

\subsubsection{Hardware Firewall}\label{sec:stud-fire}
This project~\cite{firewall} ran in parallel with the TCP project, mentioned above. The idea was that while network packets were being handled by the TCP controller, analysis of the packet headers, and possibly also packet body, could be performed in parallel.

The first solution was basic filtering based on ranges of allowed and blocked IP addresses and ports. It was implemented by having multiple filter processes, which each held their own rule. Each of these processes would output a flag, indicating whether the packet should be dropped. These flags would be collected as a single flag and sent back to the source of the packages. In this case it was the TCP controller, telling it to drop the packet, if the flag was set to \texttt{true}.

The next step would have been to implement deep packet inspection, where processes would try to match known signatures to the packet body. At the end of the project, a coarse grain architecture for solving this was presented, but not implemented due to the project deadline.

\subsubsection{Bohrium to SME}\label{sec:stud-bohsme}
In this project, BohSME~\cite{bohsme}, SME was added as a target for Bohrium~\cite{bohrium}, stretching the supported platforms of Bohrium to FPGAs.

Bohrium works as a drop-in replacement for the popular Python library Numpy. Each call to a Numpy function is captured into an AST representing the operations of the program and their order. Bohrium will not evaluate this AST until a result is needed, allowing Bohrium to make optimizations on the AST.

For BohSME, this AST is transformed into processes, handling the computations, and buses handling the transfers. The transformed program would send and receive data as streams, making it fit FPGAs more. Because the data is served as a stream, values will be presented at once, even though some might not be needed before other computations have been made. BohSME introduced registers that could hold these values until they were needed, without stopping the pipeline.

The biggest problem is that Bohrium is Just-In-Time (JIT) compiled, meaning that the program is not compiled until it is needed. This results in slow running times on small data sets and improved running time on bigger data sets, where the improvements from optimizations begin to outshine this cost of compiling. This is not feasible on FPGAs, as compilation for FPGAs takes a very long time, in the orders of hours. This means that BohSME is not usable in exactly the same way as Bohrium.

\subsubsection{RISC-V processor}\label{sec:stud-risc}
This project extended a previous masters student project implementing a MIPS processor~\cite{johnsen2017mipssme}, by implementing a more modern RISC-V processor~\cite{daniel-risc}. It followed the same steps of the previous project up until implementing a single cycle processor. Due to the project deadline, pipelining the processor was not implemented.

This project was made by a Physics student, who, at the time, hadn't had any courses, or experience, related to machine architecture. As such, there was a steep learning curve in the subjects SME, FPGAs and general computer organization. The project resulted in a RISC-V processor being able to run the \texttt{.TEXT} region of a binary produced by GCC targeting RISC-V.

\subsubsection{Feed Forward Neural Network}\label{sec:stud-amira}
This project~\cite{amira} is a work in progress and it is working on converting a trained PyTorch~\cite{pytorch} model to a hardware model through SME. It focuses on inference of a feed forward neural network.

These models are usually implemented using a series of vector and matrix operations. This project follows the same strategy and as such is implementing some of these operations, such as matrix multiplication, which is showcased in Section~\ref{sec:ex-matmul}.

The series of operations is implemented as a pipeline, where each pipeline stage follows the same structure as the project described in~\ref{sec:stud-sorting}. Each pipeline stage, streams the data between a set of BRAM. This allows for multiple inputs being in the pipeline at once.

This is still a work in progress, and therefore there are no concrete results, at the time of writing. The preliminary results looks promising, as it produces the same output as the original Pytorch model, given the same input.

\section{Examples}\label{sec:ex}
This section will show a few examples implemented using SME. Code will not be shown, but is provided online~\cite{sme-examples}.

All of the examples have been compiled for the ZedBoard, hosting a Xilinx Zynq-7000 SoC XC7Z020-CLG484-1, using Xilinx Vivado 2020.1. In order to communicate with the designs, each example have been implemented with the AXI BRAM controllers for each of the top-level BRAM buses, and general AXI registers for the rest of the buses.

\begin{table}[!t]
\renewcommand{\arraystretch}{1.3}
\caption{Performance of the implemented examples. f$_{max}$ is the maximum achievied clockrate. LUT is the amount of Look-Up Table (LUT) used. FF is the amount of Flip-Flop (FF), or registers, used. BRAM is the amount of Block Random Access Memory (BRAM) used. DSP is the amount of Digital Signal Processor (DSP) used.}
\label{tab:performance}
\centering
\begin{tabular}{lrrrrr}
\hline
Name & f$_{max}$ & LUT & FF & DSP & BRAM \\
\hline
HW Counter & 50 MHz & 129 & 7 & 0 & 0 \\
AES & 15 MHz & 17162 & 161 & 0 & 0 \\
Mat Mul & 65 MHz & 932 & 895 & 12 & 3 \\
MIPS & 70 MHz & 17830 & 5991 & 0 & 4 \\
Binning & 200 MHz & 186 & 384 & 0 & 1 \\
\hline
\end{tabular}
\end{table}

\subsection{Hardware counter}\label{sec:ex-counter}
\begin{figure}
    \centering
    \scalebox{0.7}{
        \begin{tikzpicture}
            \node[toplevel] (toplevel) at (0,0) {};
            \node[empty] (bus_ctrl) at (-4.2, 0) {\texttt{Control}};
            \node[empty] (bus_led) at (4.7, 0) {\texttt{LEDs}};

            \node[process] (counter) at (0,0) {\texttt{Counter}};
            \path[draw, ->] (-3, 0) -- (counter.180) node [very near start] {\footnotesize \texttt{active}};
            \path[draw, ->] (2.5,0) -- (3.5,0) node [very near end] {\footnotesize \texttt{value[0]}};
            \path[draw, ->] (2.5,-0.5) -- (3.5,-0.5) node [very near end] {\footnotesize \texttt{value[1]}};
            \path[draw, ->] (2.5,-1) -- (3.5,-1) node [very near end] {\footnotesize \texttt{value[2]}};
            \path[draw, ->] (counter.0) -| (2.5,0) |- (3.5,-1.5) node [pos=.94] {\footnotesize \texttt{value[3]}};
        \end{tikzpicture}
    }
    \caption{The block diagram of the hardware counter example. Processes are shown as squares. The gray dashed square shows the top-level of the network. The buses have been expanded to show the fields. It has one top-level input bus, \texttt{Control}, and one top-level output bus, \texttt{LEDs}.}
    \label{fig:counter}
\end{figure}
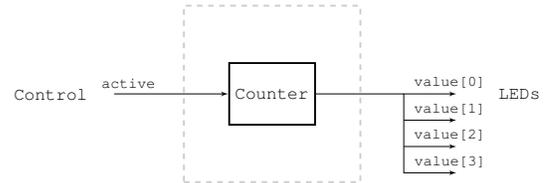

This example, as shown in Figure~\ref{fig:counter}, describes a hardware counter, that increments a counter every $n$ clock cycles. This is usually used as a toy example, since it can be implemented very easily on an FPGA. It can also provide a very visual result, if the inputs are mapped to something like a physical switch and the outputs are mapped to LED lights.

It has one input, \texttt{Control} containing the field \texttt{active}, which controls whether it should increment. It has one output, \texttt{LEDs} containing the field \texttt{value}, which is a 4 bit binary value holding the current count.

The performance of this example is not really important, as the main requirement is that it meets the given timing. In this case, we implemented it at 50 MHz, incrementing every second, so $n$ is set to 50 million. The numbers describing resource utilization can be seen in Table~\ref{tab:performance}.

\subsection{AES encryption}\label{sec:ex-aes}
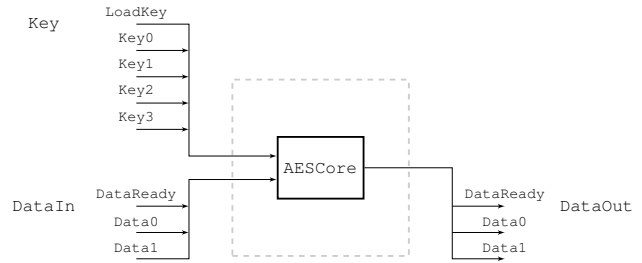
\begin{figure}
    \centering
    \scalebox{0.7}{
        \begin{tikzpicture}
            \node[toplevel] (toplevel) at (0,0) {};
            \node[empty] (bus_key) at (-5.25, 2.73) {\texttt{Key}};
            \node[empty] (bus_datain) at (-5.25, -.73) {\texttt{DataIn}};
            \node[empty] (bus_dataout) at (5.25, -0.73) {\texttt{DataOut}};

            \node[process] (aes) at (0,0) {\texttt{AESCore}};
            \path[draw, ->] (-3.5, 2.73) -| node[pos=0] {\footnotesize \texttt{LoadKey}} (-2.5,0.3) |- (aes.165);
            \path[draw, ->] (-3.5, 2.23) -- node[pos=0] {\footnotesize \texttt{Key0}} (-2.5,2.23);
            \path[draw, ->] (-3.5, 1.73) -- node[pos=0] {\footnotesize \texttt{Key1}} (-2.5,1.73);
            \path[draw, ->] (-3.5, 1.23) -- node[pos=0] {\footnotesize \texttt{Key2}} (-2.5,1.23);
            \path[draw, ->] (-3.5, 0.73) -- node[pos=0] {\footnotesize \texttt{Key3}} (-2.5,0.73);

            \path[draw, ->] (-3.5, -0.73) -- node[pos=0] {\footnotesize \texttt{DataReady}} (-2.5,-.73);
            \path[draw, ->] (-3.5, -1.23) -- node[pos=0] {\footnotesize \texttt{Data0}} (-2.5,-1.23);
            \path[draw, ->] (-3.5, -1.73) -| node[pos=0] {\footnotesize \texttt{Data1}} (-2.5,-0.3) |- (aes.195);

            \path[draw, ->] (2.5,-0.73) |- (3.5,-.73) node [pos=1] {\footnotesize \texttt{DataReady}};
            \path[draw, ->] (2.5,-1.23) -- (3.5,-1.23) node [pos=1] {\footnotesize \texttt{Data0}};
            \path[draw, ->] (aes.0) -| (2.5,-0.73) |- (3.5,-1.73) node [pos=1] {\footnotesize \texttt{Data1}};

        \end{tikzpicture}
    }
    \caption{The block diagram of the AES example. Processes are shown as squares. The gray dashed square shows the top-level of the network. The buses have been expanded to show the fields. It has two top-level input buses, \texttt{Key} and \texttt{DataIn}, and one top-level output bus, \texttt{DataOut}.}
    \label{fig:aes}
\end{figure}

This example, as shown in Figure~\ref{fig:aes}, shows how regular C\# code can be used to create hardware. It is implemented using a single process, which contains an AES implementation, taken from the Mono project~\cite{monoaessource}. For verification, the output has been compared to a call to the standard C\# library implementation in \texttt{System.Security.Cryptography.Aes}.

It has two inputs, \texttt{Key} and \texttt{DataIn}. One for loading the encryption key and the other for the data to encrypt. It has one output, \texttt{DataOut}, for the encrypted data. The \texttt{Key} bus has five fields: one \texttt{LoadKey}, indicating whether the process should store the given key, and four \texttt{Key[n]} fields holding the encryption key. The two \texttt{Data} buses has three fields: two \texttt{Data[n]}, holding the data encrypted/to be encrypted data, and one \texttt{DataReady} indicating whether the data on the two other fields are valid.

While it does map to hardware, it is not a very efficient implementation, as can be seen in Table~\ref{tab:performance}. This is mainly due to everything happening within one clock cycle, where everything could be pipelined. However, it still shows how higher level code can be implemented in hardware using SME. Given the design consumes two 64-bit words per clock cycle, the throughput is $2*64*f_{max} = 1.92$ Gbit per second.

\subsection{Matrix multiplication}\label{sec:ex-matmul}
\begin{figure*}[h!]
        \centering
        \scalebox{0.5}{
            \begin{tikzpicture}
                \node[toplevel, minimum height=18em, minimum width=48em] (toplevel) at (3.25,-0.5) {};

                \node[process] (accessgen) at (-3,-2.5) {AccessGenerator};
                \node[ram] (rama) at (-3,1.5) {RAM A};
                \node[ram] (ramb) at (-3,0) {RAM B};
                \node[ram] (ramc) at (10,-0.8) {RAM C};
                \node[process] (multiplier) at (2.5,.7) {Multiplier};
                \node[process] (accumulator) at (5.5,.7) {Accumulator};
                \node[process,minimum height=2em] (reg0) at (0,1.5) {Reg};
                \node[process,minimum height=2em] (reg1) at (0,0) {Reg};
                \node[process,minimum height=2em] (reg2) at (0,-2.5) {Reg};
                \node[process,minimum height=2em] (reg3) at (2.5,-2.5) {Reg};
                \node[process,minimum height=2em] (reg4) at (5.5,-2.5) {Reg};
                \node[process,minimum height=2em] (reg5) at (8,-2.5) {Reg};

                \node[empty] (bus_metaa) at (-7, -2) {\texttt{MatrixMetaA}};
                \node[empty] (bus_metab) at (-7, -2.5) {\texttt{MatrixMetaB}};
                \node[empty] (bus_metac) at (-7, -3) {\texttt{MatrixMetaC}};

                \path[draw, ->] (bus_metaa.0) -- (accessgen.160);
                \path[draw, ->] (bus_metab.0) -- (accessgen.180);
                \path[draw, ->] (accessgen.200) -- (bus_metac.0);

                \path[draw, ->] (accessgen.90) -- (ramb)  node[near end, left] {\footnotesize \texttt{addr}};
                \path[draw, ->] (accessgen.70) |- (-1.5, -1) -| (-1.5, 0) |- (rama.335)  node[near end] {\footnotesize \texttt{addr}};

                \node[empty] (aporta) at (-7, 1.5) {\texttt{AportA}};
                \path[draw, ->] (aporta.9) -- (rama.171);
                \path[draw, ->] (rama.190) -- (aporta.350);
                \node[empty] (bporta) at (-7, 0) {\texttt{BportA}};
                \path[draw, ->] (bporta.9) -- (ramb.170);
                \path[draw, ->] (ramb.192) -- (bporta.349);

                \path[draw, ->] (rama.0) -- (reg0.180)  node[near end] {\footnotesize \texttt{data}};
                \path[draw, ->] (ramb.0) -- (reg1.180)  node[near end] {\footnotesize \texttt{data}};
                \path[draw, ->] (accessgen.0) -- (reg2.180)  node[midway] {\footnotesize \texttt{addr}};

                \path[draw, ->] (reg0.0) -| (0.8, 1.5) |- (multiplier.170)  node[near end] {\footnotesize \texttt{data}};
                \path[draw, ->] (reg1.0) -| (0.8, 0) |- (multiplier.190)  node[near end, below] {\footnotesize \texttt{data}};
                \path[draw, ->] (reg2.0) -- (reg3.180)  node[midway] {\footnotesize \texttt{addr}};
                \path[draw, ->] (reg3.0) -- (reg4.180)  node[midway] {\footnotesize \texttt{addr}};
                \path[draw, ->] (reg4.0) -- (reg5.180)  node[midway] {\footnotesize \texttt{addr}};
                \path[draw, ->] (reg4.90) -- (accumulator.270)  node[near end] {\footnotesize \texttt{addr}};

                \path[draw, ->] (multiplier.0) -- (accumulator.180)  node[midway] {\footnotesize \texttt{data}};

                \path[draw, ->] (accumulator.0) -| (9, .7) node[midway, right] {\footnotesize \texttt{data}} |- (ramc.170)  ;
                \path[draw, ->] (reg5.0) -| (9, -2) node[midway, right] {\footnotesize \texttt{addr}} |- (ramc.190)  ;

                \node[empty] (cporta) at (12.5, -0.8) {\texttt{CportA}};
                \path[draw, ->] (ramc.9) -- (cporta.171);
                \path[draw, ->] (cporta.189) -- (ramc.350);
            \end{tikzpicture}
        }
        \caption{The block diagram of the matrix multiplication example. Processes are shown as squares. The gray dashed square shows the top-level of the network. For each BRAM, there is a set of buses exposed as top-level. There are two additional top-level input buses, \texttt{MatrixMetaA} and \texttt{MatrixMetaB}. There is one additional output bus, \texttt{MatrixMetaC}.}
        \label{fig:matmul}
    \end{figure*}
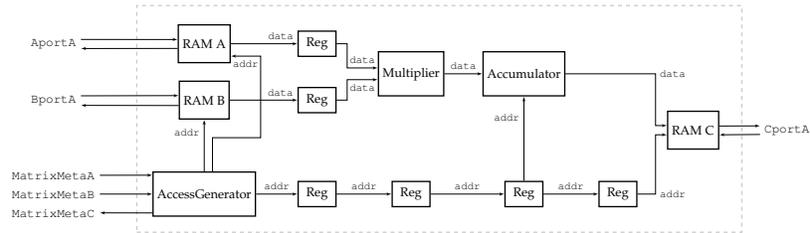

This example, as shown in Figure~\ref{fig:matmul}, shows matrix multiplication, a common High Performance Compute (HPC) problem.
It was developed as part of the masters student project described in Section~\ref{sec:stud-amira}.

It exposes the \texttt{A} port of three Block RAM holding the matrices, $A$, $B$ and $C$, used for the computation $C = A * B$. These ports are named \texttt{AportA}, \texttt{BportA} and \texttt{CportA} respectively. It has two \texttt{MatrixMeta} input buses and one \texttt{MatrixMeta} output, specifying the sizes and validity of the respective matrices. Each \texttt{MatrixMeta} bus is associated with the BRAM of the same post-fix. The bus has five fields: a \texttt{valid} indicating whether the data is valid, a \texttt{base} holding the base address, a \texttt{stride} indicating the access stride, a \texttt{height} and \texttt{width} specifying the size of the matrix.

It is implemented in a pipelined fashion, where each sub computation is performed in a single clock cycle, as is shown by the two processes \texttt{Multiplier} and \texttt{Accumulator}. On each clock cycle, one value is read from the $A$ and $B$ matrix, which are streamed to the \texttt{Multiplier}. The \texttt{Accumulator} keeps adding its input, until the address for matrix $C$ changes, where it will write the value to $C$.

The performance of the implementation can be seen in Table~\ref{tab:performance}. Each entry in $C$ takes in the order of $M*N*K$ clock cycles to compute. The main bottleneck of the implementation is the multiplication, which is performed within a single clock cycle. By pipelining it, performance should increase, with only some additional registers for transmitting the address.

\subsection{MIPS processor}\label{sec:ex-mips}
\begin{figure*}[h]
    \centering
    \scalebox{0.5}{
        \begin{tikzpicture}
            \node[toplevel, minimum height=38em, minimum width=61em] (toplevel) at (0,2.3) {};

            \node[process] (alu) at (0.75,0) {ALU};
            \node[process] (reg) at (-4,0) {Register File};
            \node[process] (memo) at (4,0) {Memory};
            \node[process] (jump) at (4,4.75) {Jump};
            \node[process] (imem) at (-8,0) {Instruction Memory};
            \node[process] (pc) at (-9, 2) {PC};
            \node[control] (cont) at (-4, 3) {Control};
            \node[mux] (wbm) at (8,1) {|};
            \node[process] (forw) at (0, -3) {Forwarding Unit};
            \node[mux] (forwa) at (-0.25, 0.3) {|};
            \node[mux] (forwb) at (-0.75, -0.3) {|};
            \node[mux] (addr) at (-4, 1) {|};
            \node[process] (haz) at (-4,6.5) {Hazard Detectection};

            \node[empty] (if) at (-8, 8) {\textbf{IF}};
            \node[empty] (id) at (-4, 8) {\textbf{ID}};
            \node[empty] (ex) at (0, 8) {\textbf{EX}};
            \node[empty] (mem) at (4, 8) {\textbf{MEM}};
            \node[empty] (wb) at (8, 8) {\textbf{WB}};

            \path[draw, ->] (pc) -| (imem);
            \path[draw, ->] (imem) -- (reg);
            \path[draw, ->] (-5.5,0) |- (cont);
            \path[draw, ->] (-5.5,0) |- (addr.220);
            \path[draw, ->] (-5.5,0) |- (addr.140);
            \path[draw, ->] (reg.16) -- (forwa);
            \path[draw, ->] (forwa) --  (alu.151);
            \path[draw, ->] (reg.344) -- (forwb);
            \path[draw, ->] (forwb) -- (alu.209);
            \path[draw, ->] (0,-0.3) |- (0,-1) -- (2.75,-1) |- (memo.210);
            \path[draw, ->] (2.5, 0) |- (wbm.150);
            \path[draw, ->] (memo) -- (6.5, 0) |- (wbm.210);
            \path[draw, ->] (wbm) -| (9.25, -1.75) -| (reg.290);
            \path[draw, ->] (alu.30) -| (1.5,1) |- (4,2) -| (jump.290);
            \path[draw, ->] (jump) |- (0,5.5) -| (pc);
            \path[draw, ->] (-8,2) |- (jump.160);
            \path[draw, ->] (-5.5,3) |- (jump.200);
            \path[draw, ->] (-1.25,-1.75) |- (forwb.220);
            \path[draw, ->] (-1.25,-1.75) |- (forwa.220);
            \path[draw, ->] (2.5, 0) |- (-1.5, -1.5) |- (forwb.140);
            \path[draw, ->] (-1.5, -1.5) |- (forwa.140);
            \path[draw, ->] (addr) -| (-3, 1.5) -| (9.5,-2) -|
            (reg.270);
            \path[draw, ->] (9.5, -2) |- (forw.353);
            \path[draw, ->] (5,1.5) |- (forw.15);
            \path[draw, ->] (-5.5,0) |- (-1.75,-1) |- (forw.170);
            \path[draw, ->] (-5.5,0) |- (-2,-1.25) |- (forw.190);

            \path[draw, ->] (cont.310) -| (alu);
            \path[draw, ->] (cont.330) -| (memo.115);
            \path[draw, ->] (cont.350) -| (wbm);
            \path[draw, ->] (cont.10) -| (9.75,-2.25) -| (reg.250);
            \path[draw, ->] (cont.30) -| (jump.250);
            \path[draw, ->] (cont) -- (addr);
            \path[draw, ->] (9.75,-2.25) |- (forw.345);
            \path[draw, ->] (5.25, 3.2) |- (forw.7);

            \path[draw, ->] (forw.114) -- (forwa);
            \path[draw, ->] (forw.142) -- (forwb);

            \path[draw, ->] (-0.75, 1.5) |- (haz); 
            \path[draw, ->] (-1, 2.37) |- (haz.350); 
            \path[draw, ->] (-0.5, 5.5) |- (haz.10); 
            \path[draw, ->] (-5.5, 4.55) -- (haz.202);
            \path[draw, -] (haz) -- (-4, 5.75);

            \node[block, minimum height=170, minimum width=10, fill=white] (ifid) at
            (-6,2.25) {};
            \node[block, minimum height=190, minimum width=10, fill=white] (idex) at
            (-2.5,1.75) {};
            \node[block, minimum height=180, minimum width=10, fill=white]
            (exmem) at (2,2) {};
            \node[block, minimum height=180, minimum width=10, fill=white]
            (memwb) at (6,2) {};

            \path[draw, ->] (-4, 5.75) -| (exmem);
            \path[draw, ->] (-4, 5.75) -| (idex);
            \path[draw, ->] (-4, 5.75) -| (ifid.92);
            \path[draw, ->] (-4, 5.75) -| (pc.115);
            \path[draw, ->] (haz) -| (ifid.88);
            \path[draw, ->] (haz) -| (pc.65);

            \node[empty] (bus_imem) at (-14, 0) {\texttt{Instruction Memory}};
            \path[draw, ->] (bus_imem.5) -- (imem.174);
            \path[draw, ->] (imem.186) -- (bus_imem.355);

            \node[empty] (bus_ctrl) at (-13, -2) {\texttt{Control}};
            \path[draw, ->] (bus_ctrl.10) -| (imem.255);
            \path[draw, ->] (imem.285) |- (bus_ctrl.350);

            \node[empty] (bus_mmem) at (4, -5) {\texttt{Main Memory}};
            \path[draw, ->] (bus_mmem.111) -- (memo.250);
            \path[draw, ->] (memo.290) -- (bus_mmem.69);
        \end{tikzpicture}
    }
    \caption{The block diagram of the pipelined MIPS processor. Original Figure from~\cite{sme-mips}. Processes are shown as squares and circles. The gray dashed square shows the top-level of the network. There are two BRAM ports exposed as top-level: \texttt{Instruction Memory} and \texttt{Main Memory}. Additionally, there is a top-level \texttt{Control} bus.}
    \label{fig:hazard}
\end{figure*}
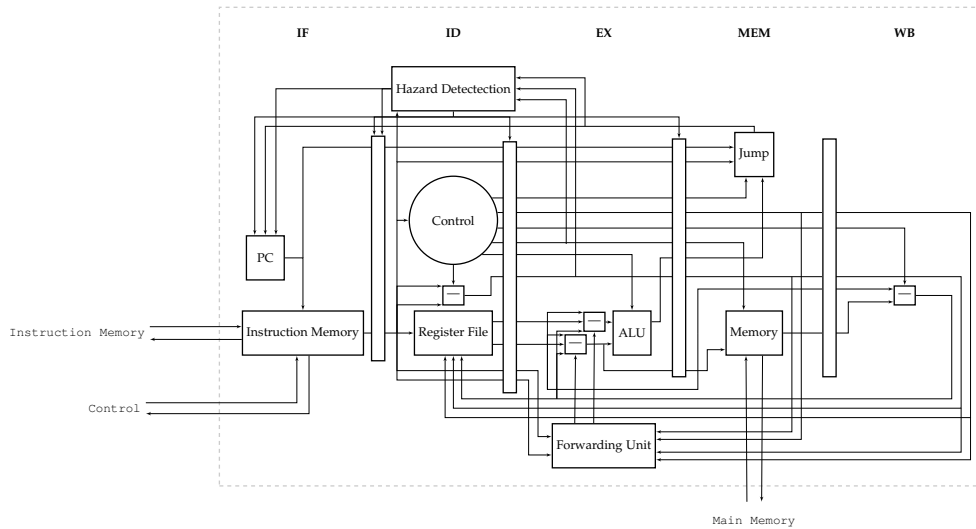

This example, as shown in Figure~\ref{fig:hazard}, shows a bigger and more complex control system in the form of the classical CPU architecture used for education: The MIPS processor. It was implemented as a masters student project, which followed the steps of a machine architecture class taught at the University of Copenhagen~\cite{johnsen2017mipssme}. The resulting design features a pipelined processor, along with logic handling the data hazards introduced by pipelining. It is not implemented using BRAM, as components weren't available at the time of writing. This example shows that controller systems, where tight control is a requirement, is a great fit for SME.

It exposes two memory ports for instruction and main memory respectively. It also features a \texttt{Control} bus for starting/stopping the processor.

It has been verified by taking programs generated by a MIPS compiler,
injecting them into the instruction memory, running the programs and verifying that the memory in the main memory is as expected. The performance of the implementation can be seen in Table~\ref{tab:performance}. The main bottleneck of the system lies within the \texttt{ALU}, as it can perform a multiplication and a division, both of which are done within a single clock cycle.

\subsection{Binning into a histogram}\label{sec:ex-maxiv}
\begin{figure}
        \centering
        \scalebox{0.7}{
            \begin{tikzpicture}
                \node[toplevel, minimum height=26em, minimum width=17em] (toplevel) at (0,-0.7) {};

                \node[process] (control) at (0,2.5) {Control};
                \node[ram] (ram) at (0,0.5) {RAM};
                \node[unclocked] (forward) at (0,-1) {Forward};
                \node[process] (adder) at (0,-2.5) {Adder};
                \node[process] (pipe) at (0,-4) {Reg/pipe};

                \node[empty] (mem) at (-4.5, 2.5) {\texttt{Mem}};
                \path[draw, ->] (mem.13) -- (control.170);
                \path[draw, ->] (control.191) -- (mem.346);

                \path[draw, ->] (ram.118) -- (control.245) node[near start] {\footnotesize \texttt{data}};
                \path[draw, ->] (control.290) -- (ram.67) node[near start] {\footnotesize \texttt{data}};

                \path[draw, ->] (control.0) -| (1.2, 2.5) |- (forward.20) node[very near start] {\footnotesize \texttt{data}};

                \path[draw, ->] (adder.155) -| (-1.2, -2.2) |- (ram.165) node[near start] {\footnotesize \texttt{data}};
                \path[draw, ->] (-1.2, -1) -- (forward.180);

                \node[empty] (value) at (-4.5, -2.5) {\texttt{Value}};
                \path[draw, ->] (value) -- (adder);

                \node[empty] (index) at (-4.5, -4) {\texttt{Index}};
                \path[draw, ->] (index) -- (pipe);

                \path[draw, ->] (pipe.0) -| (2, -4) |- (ram.0) node[near start, right] {\footnotesize \texttt{data}};
                \path[draw, ->] (pipe.0) -| (2, -4) |- (forward.0);

                \path[draw, ->] (forward.340) -| (1.2, -2) |- (adder.0) node[near start, right] {\footnotesize \texttt{data}};

            \end{tikzpicture}
        }
        \caption{The block diagram of the histogram binning example. Clocked processes are shown as solid squares and unclocked as dashed squares. The gray dashed square shows the top-level of the network. The memory port \texttt{Mem} is exposed to read the result and to reset the internal state. Inputs to the network are the \texttt{Index} and \texttt{Value} buses.}
        \label{fig:maxiv}
    \end{figure}
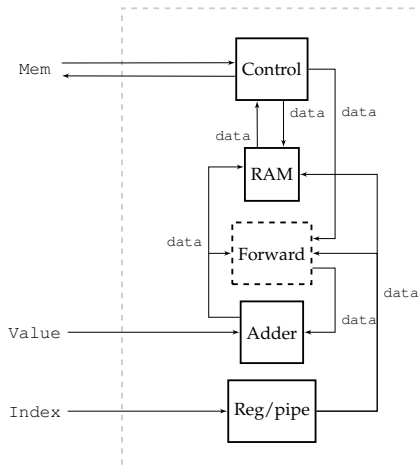

This example, as shown in Figure~\ref{fig:maxiv}, stems from a simple problem encountered at the MAX IV synchrotron in Lund, Sweden. At MAX IV, the detectors generate a massive amount of data in the orders of 100s of Gbit per second~\cite{max-iv-detektor}, when running the experiments. In many cases, they are unable to get a real-time feedback from the system while running. Even though a simple representation, in the form of a histogram, is usually sufficient. This is not feasible to do in CPU or GPU, due to the memory hierarchy of machines introducing too much overhead, but can be easily expressed as a streaming problem, as shown in this example.

The implementation has two input buses, \texttt{Index} and \texttt{Value}, indicating the index into the histogram and the value to add to that index. It also features a BRAM port, \texttt{Mem}, for resetting the internal state and reading out the result.

There exists a data hazard, when two consecutive sets of index/value have the same index. In this case, a write will be in-flight to the memory, while a read is issued, both to the same address. If this isn't handled, the value read will be the old value, not the new, resulting in incorrect results. This has been handled by the \texttt{Forwarder} process, which detects this clash and forwards the in-flight value instead of the value read from BRAM.

The performance of the implementation can be seen in Table~\ref{tab:performance}. Given the design consumes one \texttt{Index} and one \texttt{Value} per clock cycle, the system has a throughput of $2*32*f_{max} = 12.8$ Gbit per second.

\section{Discussion}
In this paper, we have iterated on the concepts of the Synchronous Message Exchange programming model in Section~\ref{sec:sme}, demonstrated by the student projects in Section~\ref{sec:stud}. They were able to implement:
\begin{itemize}
    \item A Median filter handling complex memory access patterns and sorting of data.
    \item A TCP Controller for external communication.
    \item A Firewall for filtering internet packages.
    \item BohSME, translating vector operations into SME.
    \item An implementation of a RISC-V processor.
    \item A translation of a PyTorch model into SME.
\end{itemize}
The student projects were all performed by masters students, with no prior knowledge of neither SME or FPGAs. This is a testament to the use of SME as an educational tool for learning how to design hardware models.

While SME is a higher level model compared to HDL, it is still low level compared to HLS. As a result, a user of SME still has to think in hardware structures when programming, otherwise performance will suffer. This was shown with the examples presented in Section~\ref{sec:ex}. The performance hit is shown in the AES example in Section~\ref{sec:ex-aes}. However, the fine granularity of the model is a clear fit for systems with tight requirements to when certain events happen, such as the MIPS example in Section~\ref{sec:ex-mips}. For simple problems, it is possible to achieve quite good performance, as shown in the MAXIV example in Section~\ref{sec:ex-maxiv}.

We have stated quality of life improvements to SME in Section~\ref{sec:back}. These improvements are:
\begin{itemize}
    \item Moving from .NET Framework to .NET Core, improving the portability and run-time performance of SME.
    \item Exchanging the decompiler for a compiler, to generate ASTs, and as such code, closer to the input source code.
    \item Introducing dynamic instantiation of processes and buses, to allow for multiple instances of the same entities.
    \item The addition of an intermediate language, to prevent rewriting the same optimizations for multiple front-ends.
    \item Applying formal verification to SME networks to decrease the probability of errors in the final model.
\end{itemize}
These improvements are all welcome additions to the usability of SME.

\section{Future work}
\subsection{Asynchronous semantic constructs}
We would like to leverage the asynchronous programming language construct, which have become popular in most high-level languages. In C\#, this is done by supporting \texttt{Task} and the \texttt{async} keyword. These are already utilized when simulating an SME network, but are currently unavailable inside common processes. We think these constructs can be used to express the concurrent nature of hardware models much better.

\subsection{Multi cycle buses}
In most high clocked designs, streaming data is the go-to strategy. To ensure a constant flow of data, there must not be any back-pressure from a network. A usual fix for this problem is the use of FIFO queues. We would like to implement \textit{multi-cycle buses}, which would reflect the behaviour of these queues in a more elegant matter in the C\# simulation.

\subsection{Additional components}
As has been very popular with many high-level languages, we would like to implement more library functions in the form of SME components, described in~\ref{sec:sme-components}. Examples of these are a Math library containing many common implementations of Math operations, a floating point library for introducing floating point support, and an AXI library for easy-to-use implementations of the popular AXI protocol~\cite{amba-axi}.

\subsection{High-level framework integration}
We would like to have a tighter integration into the high-level frameworks of the major vendors, as these provide a lot of helpful functionality, such as communication with other systems and hardware controllers. Both of the major vendors, Xilinx and Intel, support using Register Transfer Level (RTL) kernels as part of their respective high-level frameworks. Since SME targets VHDL, which is an RTL language, it should be possible to conform to the standard of the two vendors.

\subsection{Fully parallel execution}
As stated in Section~\ref{sec:back-cs-dotnet}, there is a false dependency when running the SME model. We believe this is due to .NET locking the buses. However, given that the SME model does not allow for multi-drivers, the dependency of multiple writers should not be present. A better approach would be locking the individual fields, rather than the entire bus. The dependency of two clocked processes reading and writing to the same bus is also false, since each process will only have one end of the bus. This should be circumvented by splitting the object into a reading and a writing end, thus locking individual ends.

\section{Conclusion}
We have presented SME, which is a useful tool for software programmers wanting to develop and test low-level hardware models. It has received a series of updates improving its usability, as presented in this paper. It requires no prior knowledge of hardware design in order to implement a working hardware model, which is shown through the student projects.

From our experience with different projects using SME, it is not a solution for everything, but there is a lot of potential for solving specific problems using SME.

\appendices

\bibliographystyle{IEEEtran}
\bibliography{biblio.bib}

\end{document}